\documentclass{aa}  
\usepackage{graphicx}
\usepackage[authoryear]{natbib}
\usepackage{txfonts}
\def\vrm{{\rm v}}
\def\kms{{\rm\,km\,s^{-1}}}
\newcount\fignum
\def\figno{\global \advance\fignum by 1 \the\fignum}
\def\thefigno{\the\fignum}
\def\figname#1{\global\advance\fignum by 1%
\xdef#1{\the\fignum}%
\the\fignum%
\relax\immediate\write1{\def\string#1{#1}}}
\font\fivebmi=cmmib6
\font\sixbmi=cmmib6	\skewchar\sixbmi='177
\font\ninebmi=cmmib10 at 9pt 	\skewchar\ninebmi='177
\newfam\bmifam
\textfont\bmifam=\ninebmi
\scriptfont\bmifam=\sixbmi
\scriptscriptfont\bmifam=\fivebmi

\mathchardef\alpha="710B
\mathchardef\beta="710C
\mathchardef\gamma="710D
\mathchardef\delta="710E
\mathchardef\epsilon="710F
\mathchardef\zeta="7110
\mathchardef\eta="7111
\mathchardef\theta="7112
\mathchardef\iota="7113
\mathchardef\kappa="7114
\mathchardef\lambda="7115
\mathchardef\mu="7116
\mathchardef\nu="7117
\mathchardef\xi="7118
\mathchardef\pi="7119
\mathchardef\rho="711A
\mathchardef\sigma="711B
\mathchardef\tau="711C
\mathchardef\upsilon="711D
\mathchardef\phi="711E
\mathchardef\chi="711F
\mathchardef\psi="7120
\mathchardef\omega="7121
\mathchardef\varepsilon="7122
\mathchardef\vartheta="7123
\mathchardef\varpi="7124
\mathchardef\varrho="7125
\mathchardef\varsigma="7126
\mathchardef\varphi="7127
 
\def\i{\relax\ifmmode{\rm i}\else\char16\fi}

\def\spose#1{\hbox to 0pt{#1\hss}}
\def\lta{\mathrel{\spose{\lower 3pt\hbox{$\mathchar"218$}}
     \raise 2.0pt\hbox{$\mathchar"13C$}}}
\def\gta{\mathrel{\spose{\lower 3pt\hbox{$\mathchar"218$}}
     \raise 2.0pt\hbox{$\mathchar"13E$}}}
\def\s#1{\widetilde{#1}}
\def\=#1{\overline{#1}}

\def\frac#1#2{{#1 \over #2}}
\def\sfrac#1#2{\leavevmode\kern.1em
  \raise.5ex\hbox{\the\scriptfont0 #1}\kern-.1em
  /\kern-.15em\lower.25ex\hbox{\the\scriptfont0 #2}}

\def\dddot#1{\ddot#1\kern-1.4pt\dot{\phantom{#1}}\kern-3pt}

\def\etal{{\it et al.\ }}

\def\={\overline}
\def\s{\ifmmode \widetilde \else \j\fi} %produces tilde in mathmode or
\def\deg{^\circ}             %for angular measure in degrees

\def\cm{{\rm\,cm}}

\def\kms{{\rm\,km\,s^{-1}}}

\def\kpc{{\rm\,kpc}}
\def\mpc{{\rm\,Mpc}}

\def\msun{{\rm\,M_\odot}}

\def\ergpscm{{\rm\,erg\,s}^{-1}{\cm}^{-2}}

\begin{document}
   \title{The Kinematics of Intracluster Planetary Nebulae and the \\
    On-Going Subcluster Merger in the Coma Cluster Core\thanks{Based
    on data collected with the FOCAS spectrograph at the Subaru
    Telescope, which is operated by the National Astronomical
    Observatory of Japan, during observing run S04A-024.}}

   \subtitle{}

   \author{Ortwin Gerhard\inst{1}, 
        Magda Arnaboldi\inst{2,3},  
        Kenneth C. Freeman\inst{4},
        Sadanori Okamura\inst{5}, \\
        Nobunari Kashikawa\inst{6}, 
        Naoki Yasuda\inst{7}} 

\institute{Max-Planck-Institut fur Extraterrestrische Physik,
  Giessenbachstrasse, D-85741 Garching, \email{gerhard@exgal.mpe.mpg.de} 
\and ESO, Karl-Schwarzschild-Str. 2, D-85748 Garching, 
  \email{marnabol@eso.org}
\and INAF, Oss. Astr. di Torino, Strada Osservatorio 20, 10025 
        Pino Torinese, Italy 
\and RSAA, Mt. Stromlo Observatory, Cotter Road, Weston Creek, ACT 
        2611, Australia, \email{kcf@mso.anu.edu.au} 
\and Dept. of Astronomy and RESCEU, School of Science, The
       Univ. of Tokyo, Tokyo 113-0033, Japan, \\
  \email{okamura@astron.s.u-tokyo.ac.jp} 
\and NAOJ, 2-21-1 Osawa, Mitaka, Tokyo, 181-8588, Japan,
  \email{kashik@zone.mtk.nao.ac.jp} 
\and Institute for Cosmic Ray Research, Univ. of Tokyo,
        Kashiwa, Chiba 277-8582, Japan,
  \email{yasuda@icrr.u-tokyo.ac.jp}}

\date{Received Oct 2, 2006; accepted March 7, 2007}

% \abstract{}{}{}{}{} 
% 5 {} token are mandatory
 
  \abstract{} {The Coma cluster is the richest and most compact of the
nearby clusters, yet there is growing evidence that its formation is
still on-going. A sensitive probe of this evolution is the dynamics of
intracluster stars, which are unbound from galaxies while the cluster
forms, according to cosmological simulations. } {With a new multi-slit
imaging spectroscopy technique pioneered at the 8.2 m Subaru telescope
and FOCAS, we have detected and measured the line-of-sight velocities
of  37 
intracluster planetary nebulae associated with the diffuse stellar
population of stars in the Coma cluster core, at 100 Mpc distance.}
{We detect clear velocity substructures within a 6 arcmin diameter
field. A substructure is present at $\sim 5000$ $\kms$, probably from
in-fall of a galaxy group, while the main intracluster stellar
component is centered around $\sim 6500$ $\kms$, $\sim 700$ $\kms$
offset from the nearby cD galaxy NGC 4874. The kinematics and
morphology of the intracluster stars show that the cluster core is in
a highly dynamically evolving state. In combination with galaxy
redshift and X-ray data this argues strongly that the cluster is
currently in the midst of a subcluster merger, where the NGC 4874
subcluster core may still be self-bound, while the NGC 4889 subcluster
core has probably dissolved. The NGC 4889 subcluster is likely to have
fallen into Coma from the eastern A2199 filament, in a direction
nearly in the plane of the sky, meeting the NGC 4874 subcluster
arriving from the west.  The two inner subcluster cores are presently
beyond their first and second close passage, during which the
elongated distribution of diffuse light has been created. We predict
the kinematic signature expected in this scenario, and argue that the
extended western X-ray arc recently discovered traces the arc shock
generated by the collision between the two subcluster gas halos. Any
preexisting cooling core region would have been heated by the
subcluster collision.} {}


\titlerunning{Kinematic Substructures in the Coma Cluster Core}
\authorrunning{O.~Gerhard et al. }

   \maketitle
%
%________________________________________________________________

\section{Introduction}

Diffuse intracluster light (ICL) has now been observed in nearby
\citep{Feldmeier04,Mihos05} and in intermediate redshift clusters
\citep{Zibetti05,Krick06}. Individual intracluster stars have been
detected in the Virgo and Coma clusters
\citep{Ferguson98,Arnaboldi03,Gerhard05}. Recent studies show that
the intracluster light contains of the order of 10\% of the mass in
stars overall \citep{Aguerri05,Zibetti05}, but in cores of dense and
rich clusters like Coma, the local ICL fraction can be as high as
40-50\% \citep{Thuan77,Bernstein95}.

The large scale structure of the ICL in nearby clusters is rather
complex. The recent surface brightness measurements of the ICL in the
Virgo cluster by \citet{Mihos05},  down to $\mu_V=28 {\rm mag
\, arcsec}^{-2}$, have shown that the ICL is made up of a wealth of
diffuse features ranging from extended low surface brightness
envelopes around giant ellipticals to long, thin streamers, as well as
smaller scale features associated with many Virgo galaxies. A similarly
deep wide-field image  of the ICL in the Coma cluster is not yet
available, but existing observations indicate similar complexity on a
variety of scales, including tidal features like streamers
\citep{Gregg98} or arcs \citep{Trentham98}, extended halos around D
galaxies \citep{Baum86,Adami05a} and extended background light
\citep{Melnick77,Thuan77}.

Morphological studies of the ICL have some limitations though: they
cannot establish whether the ICL is made up by diffuse halos
superposed along the LOS, but still physically bound to the galaxies
they surround, or rather by stars that are free-flying in the cluster
potential.  This question must be answered by spectroscopic
observations of the line-of-sight velocities of intracluster stars.
Presently, the only tracers which allow us to measure the {\it
kinematics} and {\it dynamics} of the ICL are Intracluster planetary
nebulae (ICPNe), whose [OIII]5007\AA\ emission can be used both for
their identification and radial velocity measurement.

By measuring the projected phase space of ICPNe we can constrain the
dynamical age of the ICL component, how and when this light originated
\citep{Napolitano03,Murante04,Willman04,Murante07}. Owing to the small
fluxes of distant ICPNe this was so far only possible for the nearby
Virgo cluster \citep{Arnaboldi04}.  With a new multi-slit imaging
spectroscopy (MSIS) technique, essentially a spectroscopic blind
search technique, we have now been able to measure ICPN velocities at
substantially fainter fluxes. In our first application of this
technique we have measured the velocity distribution of ICPNe in a
field in the inner core of the Coma cluster, approximately at the peak
of the X-ray emission \citep{Gerhard05, Arnaboldi07}.

The Coma cluster (A1656) is the richest and most compact of the nearby
clusters, and has been the subject of extensive study.  Originally
thought of as the prototype of a rich, regular, and relaxed cluster,
it is now known to contain significant small-scale and large-scale
substructure in the galaxy distribution
\citep{Fitchett87,Mellier88,Biviano96}, galaxy velocity distribution
\citep{Colless96,Adami05b}, and X-ray emissivity
\citep{Briel92,White93,Neumann01,Neumann03}.  Based on the recent work
Coma is believed to be undergoing several accretion and merger
events. Notably, the subcluster around NGC 4839 can be well
distinguished in both the galaxy distribution and velocities and in
the X-ray maps, and is falling into the main Coma cluster from the SW,
at a velocity of $\sim 1700\kms$
\citep{Mellier88,Colless96,Neumann01}.  But also the cD galaxy NGC
4889 together with an associated subcluster are believed to be merging
with the main Coma cluster, often associated with NGC 4874, although
the details are less clear \citep{Colless96,Adami05b}. Finally,
\citet{Adami05b} list a number of substructures, many of which may be
infalling into the Coma cluster as well.

The purpose of this paper is to ask what can be learnt by combining
the measurement of the ICL kinematics in our field in the Coma core
with the properties of the galaxy velocity distribution and X-ray
morphology. Section~\ref{observ} contains a brief summary of the MSIS
technique and discusses the criteria for identifying the different
emission sources in the data. We distinguish between galaxy PNe and
ICPNe in the field, and discuss the velocity histogram of ICPNe. The
detailed procedures used for the data reduction, the catalogue of PNe,
and their spatial and magnitude-velocity distributions are given in
\citet{Arnaboldi07}. In Sect.~\ref{mainpk} we discuss the velocity
distribution of the ICPNe sample in relation with the galaxy redshift
distribution and some notable features in the X-ray emission in the
Coma cluster core, and develop a model for the on-going merger of the
NGC 4889 subcluster with the second main subcluster in Coma around NGC
4874.  Finally, we give our conclusions in Section~\ref{end}.  In what
follows we assume a distance to the Coma cluster of 95 Mpc, thus $1''
= 460 {\rm pc}$.

%__________________________________________________________________

\section{The velocity distribution of intracluster stars in a
Coma core field from MSIS Observations}\label{observ}

\subsection{The MSIS technique}

Beyond about $20-30\mpc$ distance, PNe are too faint to be detectable
with narrow band surveys or slitless spectroscopy - their emission
disappears in the sky noise in the narrow band filter. The brightest
PNe in the Coma cluster at $100\mpc$ distance have line fluxes of
$2.2\times 10^{-18} \ergpscm$, corresponding to $\sim 20$ photons per
minute through an 8m telescope aperture, of which $\sim 2$ will reach
the detector for a typical $\sim 10\%$ overall system efficiency.  To
detect such distant and faint PNe requires a spectroscopic blind
search technique: spectroscopic, so that only the sky noise within a
few \AA\ dilutes the emission from the PN, and blind, because the
positions of these faint PNe cannot be previously determined.

The Multi-Slit Imaging Spectroscopy (MSIS) technique is such a
spectroscopic blind search technique. Its first application to the
detection of PNe in the Coma cluster was described by
\cite{Gerhard05}.  The technique combines a mask of parallel multiple
slits with a narrow-band filter, centered on the redshifted
[OIII]$\lambda5007$\AA\ emission line.  Spectra are obtained of all
PNe that happen to lie behind the slits. The narrow band filter limits
the length of the spectra on the CCD so that many slits can be
simultaneously exposed.  For each set of mask exposures only a
fraction of the field is surveyed; to increase the sky coverage the
mask can be stepped on the sky.  The technique is similar to the
approach used in the search for Ly$\alpha$ emitting galaxies at very
high redshifts \citep{SternSpinrad99, Tran04, MartinSawicki04}.

Our pilot survey in the Coma cluster was carried out with the FOCAS
spectrograph at the Subaru telescope.  The instrument was used
with a mask of 70 parallel slits and a narrow band filter of FWHM 60
\AA, centered on the [OIII] 5007\AA\ line at the redshift of the Coma
cluster.  The fraction of the field surveyed by a single mask is then
$\sim12\%$.  Each spectrum extends over 43 pixels on the CCD, and the
spectral resolution is $440\kms$. As these values show, a compromise
must be found in the MSIS setup between the number of slits and hence
number of objects detected, and the velocity resolution and
signal-to-noise (S/N) of the emission sources, which are dependent on
the spectral resolution. For a detailed description of the
observing technique and the signal-to-noise calculations we refer to
\citet{Gerhard05}.

Three mask configurations were observed for the Coma field centered at
$\alpha(J2000)\, 12:59:41.784$; $\delta(J2000)\, 27:53:25.388$, near
the centroid of the X-ray emission in the Coma cluster core, nearly
coincident with the field observed by \citet{Bernstein95} and $\sim5$
arcmin away from the cD galaxy NGC 4874. The data reduction of the
three masks observed with the MSIS technique was carried out in IRAF
and is described in \citet{Arnaboldi07}. The removal of the
instrumental signature was done following standard procedures for CCD
reductions.  The final dispersed image of the field resembles
a brick wall made up of adjacent 60 \AA\ wide two-dimensional spectra;
see Fig.~2 of \cite{Gerhard05} and Fig.~1 of \cite{Arnaboldi07}.
In this image point-like, monochromatic emission sources appear unresolved 
in both the spatial and wavelength directions, stars appear as spatially
unresolved 60 \AA\ continuum spectra, and galaxies appear as extended
blobs.

\subsection{Emission sources in the pilot study MSIS field}

In the final co-added frames we look for emission line objects and
classify them according to the following criteria, described in more
detail in \citet{Gerhard05} and \citet{Arnaboldi07}:

\begin{itemize}
\item unresolved (both in wavelength and in space) emission line
objects with no continuum, which are our intracluster planetary
nebulae (ICPNe) candidates;
\item spatially unresolved continuum sources with unresolved or
resolved line emissions. These are most likely background galaxies;
\item unresolved line emissions associated with the extended continuum
halos of Coma cluster galaxies in the field - such objects are
compatible with being PNe associated with the stellar population
emitting the continuum light. 
\end{itemize}

In total 60 such emission line objects were detected, of which
according to these criteria
\begin{itemize}
\item 35 PN candidates are ICPNe,  with no detectable continuum flux
down to $1.6\times 10^{-20} \ergpscm{\rm \AA}^{-1}$, and equivalent
widths ranging from EW $> 110$ \AA\ to 18 \AA; 
\item 20 are background objects;
\item 5 are PN candidates possibly associated with extended continuum
halos from nearby Coma galaxies.
\end{itemize}

The identification of the objects unresolved in both position and
velocities as PNe is further supported by the fact that the emission
fluxes of the brightest objects are consistent with those of the
brightest PNe in a population at distance 100 Mpc, and by the fact
that both lines of the [OIII] doublet were seen in all four sources
with sufficient flux  $\ge 4.0\times 10^{19} \ergpscm$
 in the [OIII] 5007 \AA\ line and at sufficiently large
recession velocity ${\rm v}_{obs} > 7400$ kms$^{-1}$ that also
$\lambda4959$\AA\ is redshifted into the wavelength range probed
\citep{Gerhard05, Arnaboldi07}. Furthermore, as we shall see below,
their distribution of recession velocities is centered around the Coma
cluster and is inconsistent with a population of background objects
uniformly distributed in velocity.  It is possible that the
sample contains a remaining contamination from so far undetected,
unresolved low-luminosity background emission line galaxies, but the
continuum limit rules out compact HII regions such as or brighter than
observed in Virgo \citep{Gerhard02}.  The final catalogue of PN
candidates from the whole dataset is given in
\citet{Arnaboldi07}.

\subsection{PNe in Coma Galaxies}

For the PN candidates superposed on the extended continuum halos of
Coma galaxies, as well as for the ICPN candidates near such halos, it
is necessary to check the measured LOS velocity against that of the
galaxy, in order to confirm or rule out the PN candidate-galaxy
association.  In the following, the IDs of the ICPN candidates refer
to their entries in the catalogue in \citet{Arnaboldi07}. 
\begin{itemize}
\item
   The measured LOS velocities for IPN217 and IPN224, $6440$ and $6364
   \kms$ respectively, are consistent with the systemic velocity of
   the galaxy NGC 4876 onto which they are superposed, for which 
   ${\rm v}_{sys} = 6678 \kms$.
\item
   The measured LOS velocity of candidate IPN114 is $6410 \kms$, not
   consistent with the systemic velocity of the underlying galaxy
   CGCG160-235, ${\rm v}_{sys} = 8114 \kms$. Similarly, for the two
   candidates IPN225 ( ${\rm v}_{\rm LOS} = 5861 \kms$) and IPN113
   (${\rm v}_{\rm LOS} = 6719 \kms$), the comparison with the systemic
   velocities of the nearby Coma galaxies [GMP83-3376]${\rm v}_{sys} =
   6815 \kms$ and [GMP83 3383]${\rm v}_{sys} = 4640 \kms$ shows very
   large differences, of the order of $1000 \kms$. These three PN
   candidates are therefore not bound to the galaxies, similarly as
   the Virgo ICPNe discovered serendipitously along the line-of-sight
   to M86 \citep{Arnaboldi96}.
\item
   However, close to IPN114 and CGCG160-235, one of the IPN candidates,
   IPN104, has ${\rm v}_{\rm LOS} = 8520 \kms$ which, given the
   spectral resolution, is similar enough to the velocity of
   CGCG160-235 that IPN104 may be bound to CGCG160-235, even though it
   is not directly superposed on the galaxy's continuum image. Halos
   extended out to $70 {\rm kpc}$ in virial equilibrium have been
   detected in Virgo ellipticals via surface brightness measurements
   \citep{Mihos05} and ICPN ${\rm v}_{\rm LOS}$ measurements
   \citep{Arnaboldi04}.

\end{itemize}
In what follows, we consider the PN candidates IPN114, IPN225 and
IPN113 as intracluster PN, while IPN217, IPN224, IPN104 are considered
as PN bound to Coma Galaxies. Thus the total number of ICPNe detected
in our Coma field is 37.

\begin{figure*}
   \centering
   \includegraphics[width=5cm]{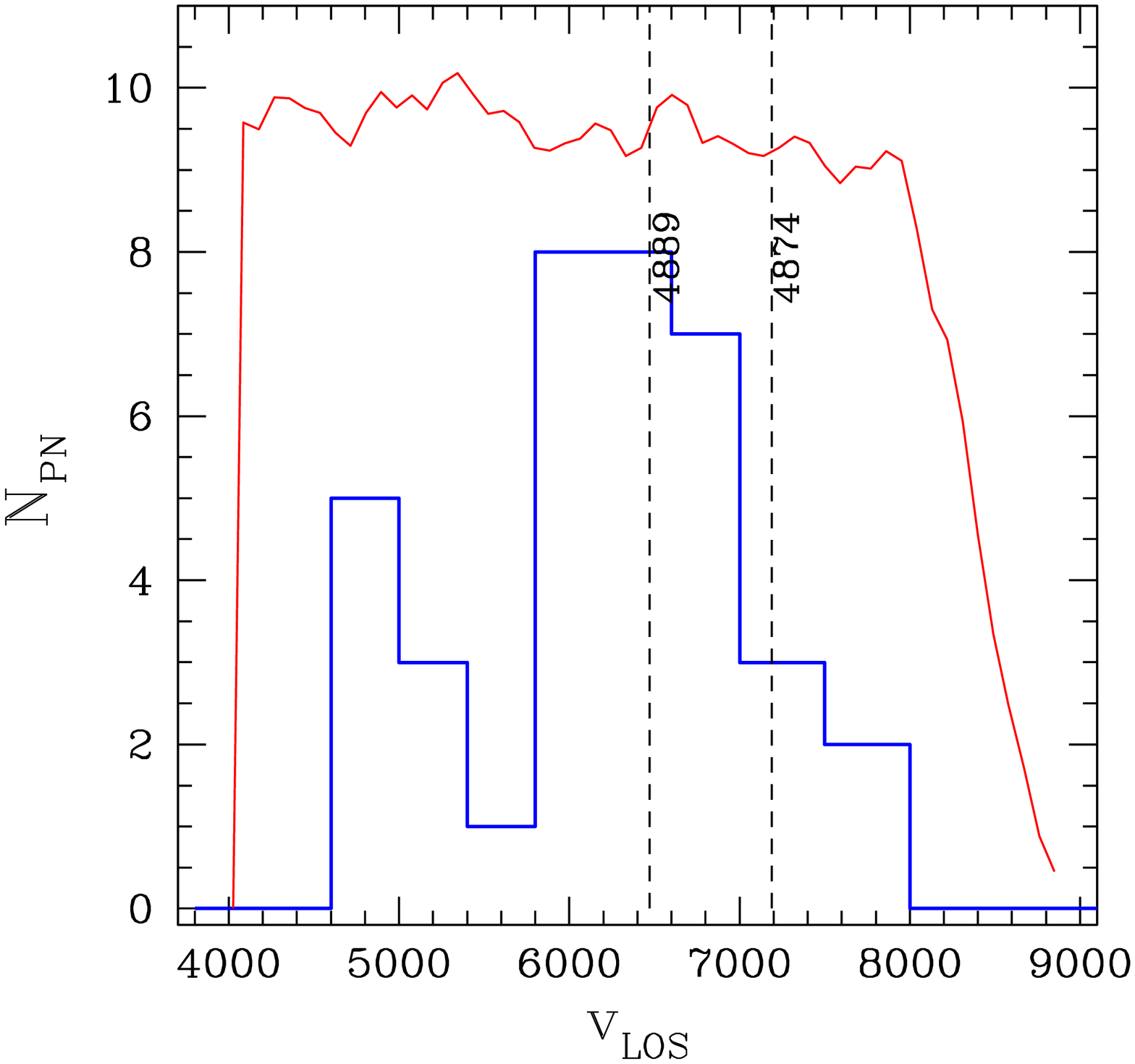}
   \includegraphics[width=5cm]{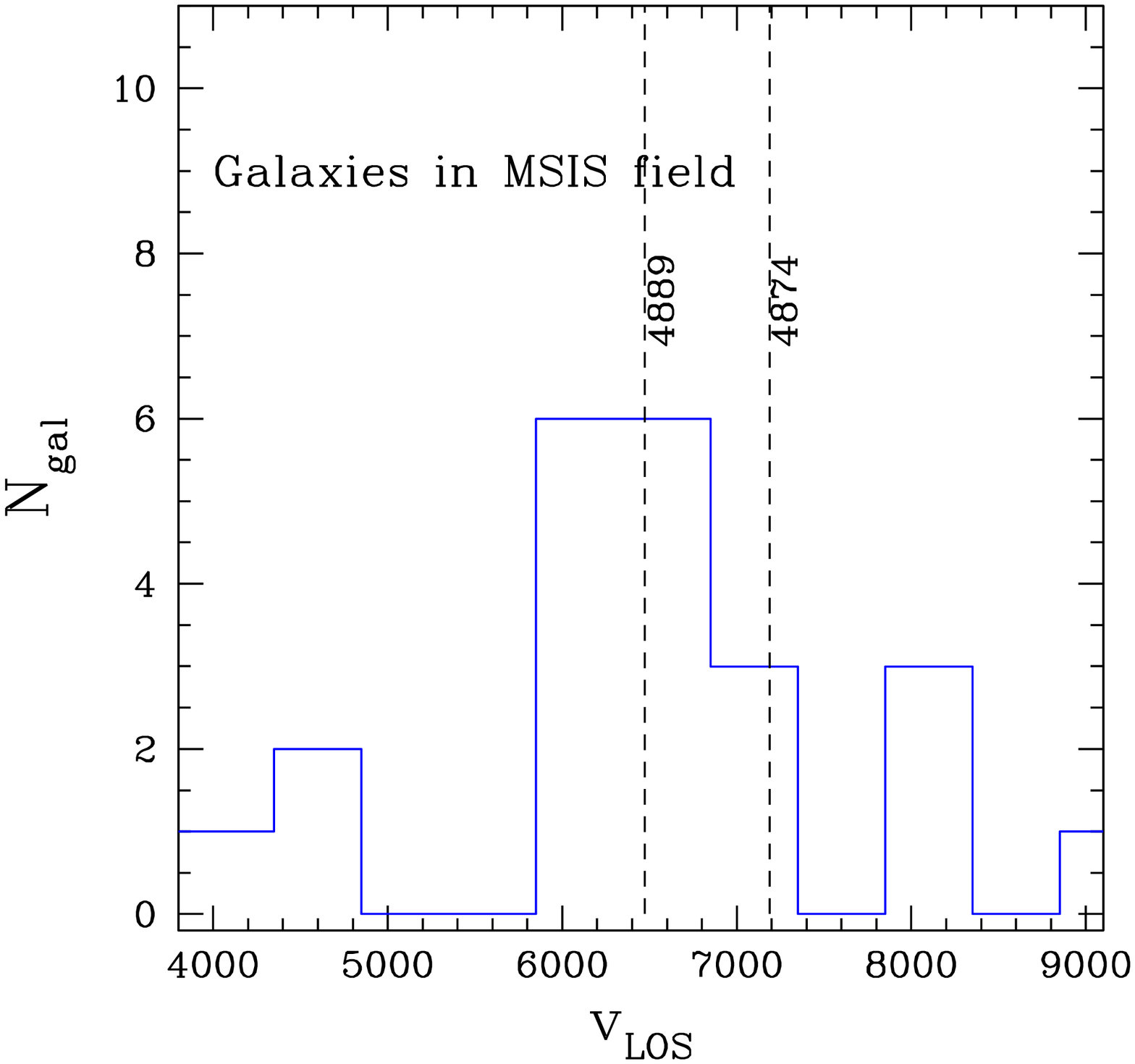}
   \includegraphics[width=5cm]{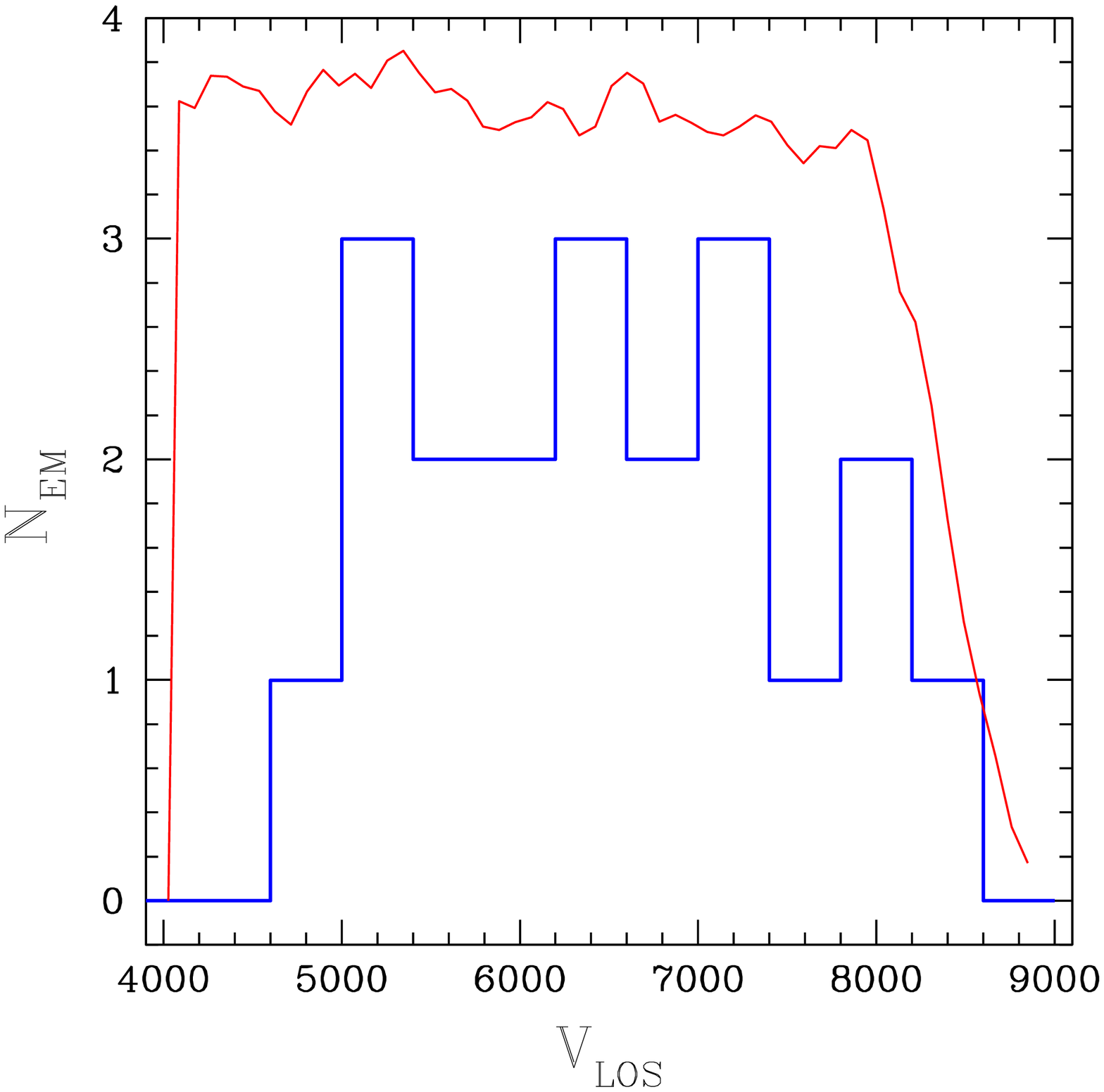}
\caption{LOS velocities of ICPNe, Coma galaxies, and background
emission galaxies in the MSIS field. Left: histogram of the LOS
velocities of ICPNe in our Coma field. The thin red line shows spectra of
 a star as measured through the MSIS slits, giving the
filter bandpass expressed in velocity.  The velocities of the
two supergiant galaxies NGC 4889 and NGC 4874 in the Coma core are
indicated by vertical dotted lines; that of the nearer NGC
4874 is displaced from the main peak of the ICPN distribution.
Center: LOS velocity histogram for the Coma galaxies in a $6' \times
6'$ field centered on the MSIS field, whose velocities would fall in
the velocity window allowed by the MSIS observing technique. The
galaxy redshifts are from \citet{Adami05b}.  Right: LOS velocity
distribution of the line emitters associated with spatially unresolved
continuum sources: these are compatible with a uniform population of
background objects. Thin red line as in the left panel.}\label{fig1}
\end{figure*}

\subsection{Velocity distribution of ICPNe}
  \label{ICPNvels}

The LOS velocity distribution of all ICPN candidates is shown in
Figure~\ref{fig1}: this shows a clear association with the Coma
cluster. The average velocity of the distribution is 6315 $\kms$ and
the standard deviation is 867 $\kms$, but two substructures are also
clearly visible: a main peak at ${\rm v}_{\rm LOS}\sim 6500 \kms$ and
a secondary peak at ${\rm v}_{\rm LOS}\sim5000 \kms$. The main peak in
the LOS velocity distribution of PN candidates is very close to the
systemic velocity of NGC 4889, ${\rm v}_{sys} = 6495 \kms$. The secondary
peak could be associated with a dissolved galaxy from one of the
subgroups discussed by \citet{Adami05b}; see Section~\ref{fiveth}
below.

Fig.~\ref{fig1} also shows the LOS velocity distribution of Coma
cluster galaxies in a $6'\times 6'$ field centered on the MSIS
pointing. Also this distribution has multiple peaks: the middle one is
coincident with the ${\rm v}_{sys}=6495 \kms$ of NGC~4889, and two
less prominent peaks are at ${\rm v}_{\rm LOS} < 5000 \kms$ and ${\rm
v}_{\rm LOS} \sim 8100 \kms$. Thus the LOS velocity distributions of
ICPNe and galaxies in the field more or less correspond; 
according to a Kolmogorov-Smirnov test, the probability of both being
drawn from the same distribution is 16\%. 

As a consistency check, we also show in Fig.~\ref{fig1} the
distribution of the observed emission lines of the background objects
within the filter passband.  For an easier comparison with the ICPN
diagram, we express the $\lambda_{\sc obs}$ in velocities, assuming
that $\lambda_0 = 5007$ \AA. The distribution of point-like
continuum sources with emission line is consistent with a flat
distribution within the filter passband window used for the MSIS
technique, and is consistent with a uniform background galaxy
population\footnote{ The apparent absence of a few objects on the blue
edge of the filter bandpass is probably caused by a slight offset
between the flat field image and the scientific image, which amplifies
the noise on this side of the spectrum.}. There is no clear
association with the Coma cluster here.  The difference to the
velocity distribution of Coma ICPNe is clear, and is one further
argument in support of our source classification.

\section{Kinematics of diffuse light and the subcluster merger in
the Coma cluster}\label{mainpk}

How does our measurement of the intracluster light kinematics in the
Coma cluster core impact on the dynamical understanding of this
densest and richest of the nearby clusters?

Optical and X-ray data have shown that the Coma cluster contains major
substructures around the infalling giant galaxy NGC 4839 as well as
around the central pair of supergiant galaxies NGC 4874 and NGC 4889
\citep{Fitchett87,Colless96,Neumann03}. In addition, a variety of
smaller substructures have been identified in position-radial velocity
data \citep{Colless96,Adami05b}. These findings, combined with the
modern understanding of structure formation in the hierarchical
Universe, have lead to the view that the cluster continues to accrete
groups from the surrounding large-scale structure, and possibly has
recently had a larger accretion of a subcluster centered around NGC
4889 . The other supergiant galaxy in the core, NGC 4874, is often
regarded as the true nucleus of the Coma cluster in these
scenarios. This is based on the fact that NGC 4874 is surrounded by
the largest concentration in the projected distribution of galaxies
\citep{Biviano96}, as well as to some extent on the fact that it has a
relatively large diffuse light halo. The fact that its radial velocity
is significantly offset relative to the velocities of nearby galaxies
is then attributed to the interaction with NGC 4889, which could have
perturbed NGC 4874 out of the cluster core \citep{Colless96}.

\subsection{The intracluster light}

\begin{figure}
   \centering
   \includegraphics[viewport=0 0 917 719,clip=true,width=\columnwidth,angle=0]{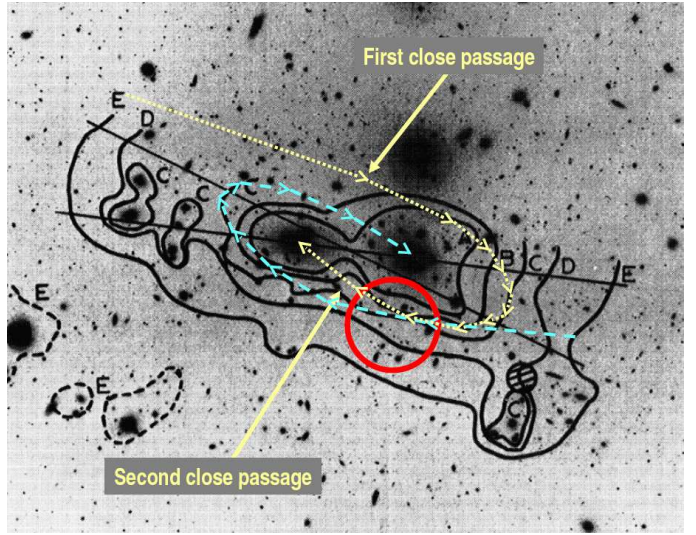}
\caption{The position of our MSIS field (red circle) on the diffuse
light isodensity contours drawn by \citet{Thuan77} in the Coma cluster
core.  The MSIS field is about 5' south of NGC 4874. The second Coma
cD galaxy NGC 4889 is 7' east  (to the left) of NGC 4874. 
 The bright
object north of NGC 4874 is the star which prevented \citet{Thuan77}
from reliably determining the northern parts of the isodensity curves
in their photographic photometry. Note the strong elongation of the
distribution of ICL in the Coma cluster core. The likely orbits of NGC
4889 and NGC 4874 up to their present positions are sketched by the
yellow dotted and magenta dashed lines, respectively; see Section
\ref{secondpassage}.  }\label{icldistr}
\end{figure}

How does the velocity distribution of ICPNe measured in our MSIS field
some $\sim 130\kpc$ from NGC 4874 fit into this picture? And how can
we use it to derive some more specific constraints on the merger
history of the Coma cluster?  The main peak of this velocity
distribution is centered not at the systemic velocity of NGC 4874 at
$\vrm = 7224\kms$, but around $\vrm = 6500\kms$, some $700\kms$ off
and exactly at the systemic velocity of the other, more distant
supergiant galaxy NGC 4889.  Clearly, these ICPNe are not bound to NGC
4874 itself, the velocity dispersion of this cD galaxy is $284\kms$
\citep{Smith00}, although they might be bound to the subcluster core
around NGC 4874 if this is still present.

Figure \ref{icldistr} shows the position of our MSIS field relative to
the two supergiant galaxies and the isodensity contours of ICL drawn
by \citet{Thuan77}. These early ICL measurements were photographic,
but agreed well with previous photoelectric measurements by
\citet{Melnick77}. The contours are well-defined south of the two
cluster dominant galaxies, while there are significant uncertainties
on the northern side due to stray light from a bright star. In
projection both galaxies appear to be embedded in a common envelope of
diffuse light; as Thuan \& Kormendy state, the light is not associated
with any particular galaxy. The large elongation of this diffuse light
distribution together with our kinematic measurement make a strong
case that this ICL in the core of Coma is in a highly dynamically evolving
state
-- we clearly do not see a relaxed ellipsoidal distribution of IC
stars here. Moreover, the paucity of diffuse light stars in this field
with NGC 4874 velocities shows that the ICL in this area is poorly
mixed. Mixing is expected to change the velocity distribution over a
typical orbital time-scale, so the ICL in this area must have been
released very recently.

\subsection{The subcluster cores}
\label{subcores}

\begin{figure}
\setlength{\unitlength}{0.01\columnwidth}
\begin{picture}(100,84)
 \put(1,1)  {\includegraphics[width=98\unitlength,bb=18 204 592 685,clip]{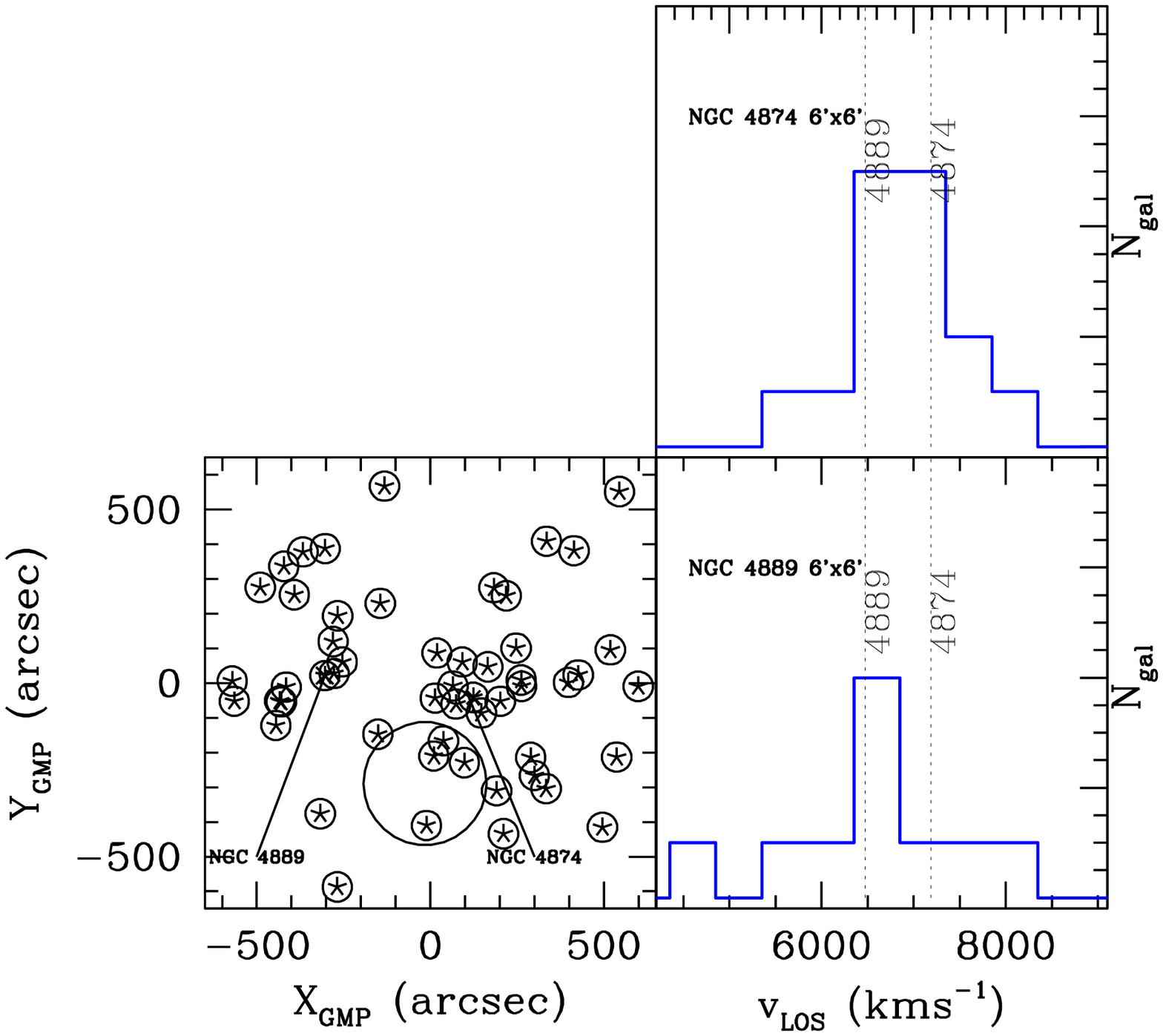}}
 \put(21,47) {\includegraphics[width=36\unitlength]{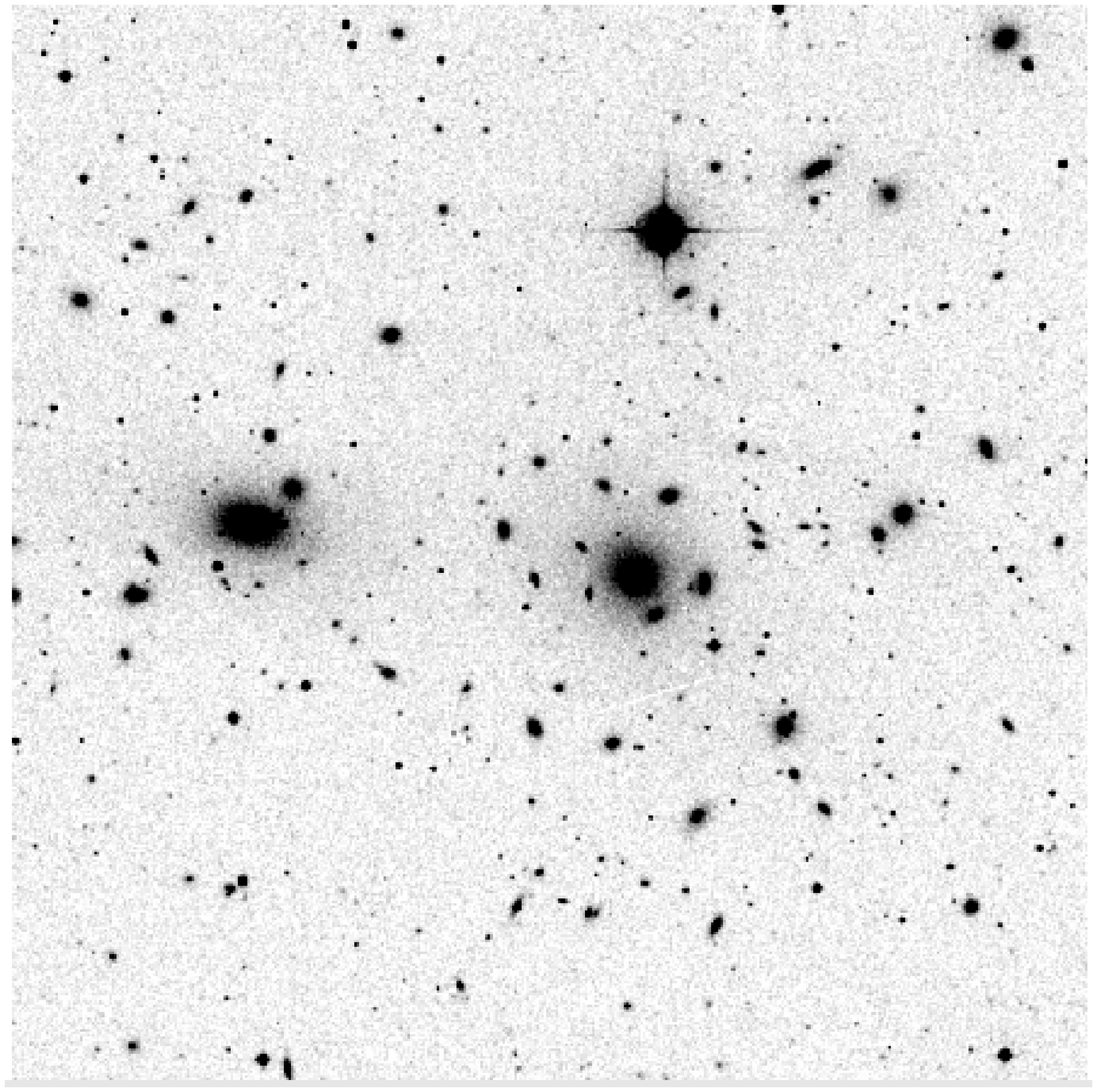}}
\end{picture}
\caption{The two subcluster cores around NGC 4874 and NGC 4889.
Top-left: DSS image of the Coma cluster center. Bottom-left: positions
of the galaxies in this region brighter than V=17; these are
concentrated around NGC 4874 and NGC 4889.  Right: velocity
distributions for these bright galaxies that fall within a 36 square 
arcmin region around NGC 4874 (top) and NGC 4889 (bottom).
}\label{subcluster}
\end{figure}

Then the galaxies and subcluster cores from which the ICL is likely to
derive will be in a state of strong interaction as
well. \citet{Korchagin01} consider a head-on collision of the two cD
galaxies and suggest that this could give rise to the elongated ICL
distribution, as could a collision with massive cluster
substructures. Another illustration is the early binary model
of \citet{Valtonen79}, who estimated the masses of the subcluster
cores around NGC 4874 and NGC 4889 and attempted to understand the
dynamics of the Coma cluster in terms of the effects of this binary on
the surrounding galaxies. Later work by \citet{Fitchett87} and
\citet{Biviano96} has confirmed the existence of subclumps in the
galaxy distribution centered on the two supergiant galaxies.  The two
dominant galaxies are projected onto the two mean peaks of the galaxy
distribution. However, contrary to naive expectation, their velocities
do {\it not} match the mean velocities of the peaks they are projected
onto.  As \citet{Colless96} state: ''the positions of the peaks and
the dominant galaxies appear to be almost mirror-symmetric in
velocity''.

However, this may be caused to considerable extent by the projection
of foreground and background galaxies onto the subcluster cores. Such
projection effects can be reduced by concentrating only on the
brightest galaxies, which are likely to be physically closer to the
dominant galaxies NGC 4874 and NGC 4889 \citep{Valtonen79,Biviano96}.
The distributions and kinematics of the bright galaxies around the two
supergiant galaxies are illustrated in Figure~\ref{subcluster}, which
is based on the catalogue of \citet{Biviano96}. The top-left part of
the figure shows a DSS image of the cluster center. In the bottom-left
panel we show the positions of the galaxies brighter than V=17 in the
region.  Their concentration around NGC 4874 and NGC 4889 is apparent;
a more quantitative analysis is in the references given above and will
not be pursued here. The remaining two panels show the velocity
distributions of these bright galaxies that fall in 6 arcmin squared
regions around NGC 4874 and NGC 4889, respectively. The two histograms
show an interesting difference: while the distribution of velocities
around NGC 4874 is compact and may indicate a still self-bound cluster
core, that around NGC 4889 is clearly not compact and has a twice
larger velocity width.  The NGC 4889 subcluster core may thus
currently be dissolving, consistent with the fact that the diffuse
light in our MSIS field 8' west of NGC 4889 may also have been tidally
dissolved from the surroundings of NGC 4889, based on its similar
radial velocity.

\subsection{The NGC 4889 subcluster orbit}

   \begin{figure}
   \centering
   \includegraphics[width=\hsize]{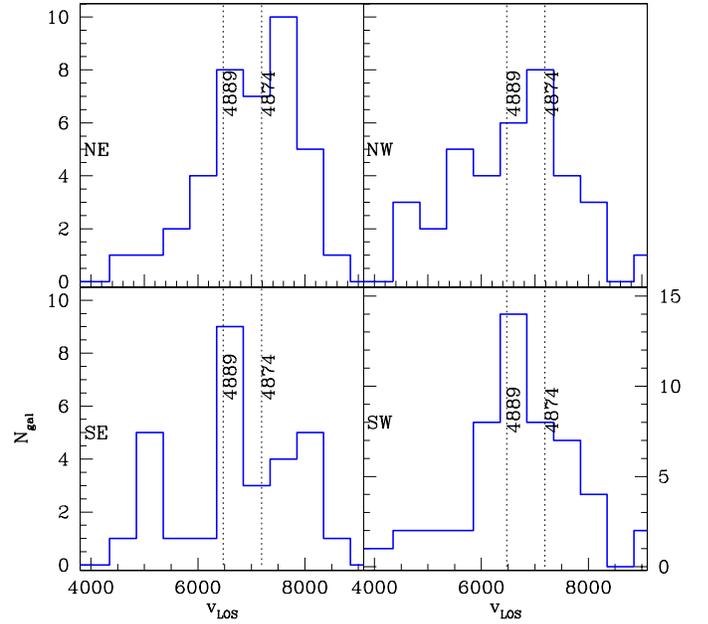}
\caption{ The ${\rm v}_{\rm LOS}$ distributions of Coma galaxies in 4
  quadrants relative to the GMP83 Coma cluster center.  Given in
  arcsec intervals along the RA and DEC directions,    NE is [-600:0,
  0:600], NW is [0:600, 0:600], SE is [-600:0,-600,0], SW is
  [0:600,-600:0]. In GMP83 coordinates NGC 4889 is at [-304,22], i.e.,
  in the NE quadrant, and NGC 4874 is at [124,-41], i.e., in the SW
  quadrant. The MSIS field is in the SE, at
  [-15,-289].}\label{quadranti}
\end{figure}

Figure \ref{quadranti} shows the radial velocity histograms for all
Coma galaxies from the catalogue of \citet{Adami05b}, in four
quadrants of size (10 arcmin)$^2$ NE, NW, SW, SE of the cluster center
defined by \citet[][hereafter GMP83]{Godwin83}.  The strong variations
from one histogram to the next show that the whole Coma cluster core
is far from complete virialization. Note, for example, the strong
peaks in the SE quadrant at ${\rm v_{\rm LOS}}=5000 \kms$ and at $6500
\kms$; our MSIS field lies in the SE/SW quadrants
(Fig.~\ref{subcluster}).  However, the real surprise of
Fig.~\ref{quadranti} is that it shows clear velocity peaks around the
LOS velocity of NGC 4889 in the SE and SW quadrants, with a further
less significant peak in the NE quadrant, and clear peaks around the
velocity of NGC 4874 in the NE and NW quadrants.  A similar
diagram showing the velocity histograms in (20 arcmin)$^2$ sized
quadrants shows a clear peak in the NE and a very strong peak in the
SW around the NGC 4889 velocity, of which the SW peak is also robust
against excluding the galaxies in the cluster center. Also, this
diagram shows robust peaks around the NGC 4874 systemic velocity in
the NE and NW. 

This is contrary to expectation because in a general
subcluster merger, these velocity distributions should have been
changed and shifted by the gravitational pull during the interaction.
The only reasonable explanation for the histograms in
Fig.~\ref{quadranti} is that the average orbital accelerations on the
galaxies of the two interacting subclusters on scales of $\sim
500\kpc$ have predominantly been in the plane of the sky. Obviously
this is a lucky coincidence; for a general orbit, the line-of-sight
component of the acceleration would be visible in the distribution of
galaxy recession velocities.  For example, in a head-on collision
along the line-of-sight, the observed galaxy velocities at the time of
strong interaction would have been modified most in the interacting
cluster cores and least in the subcluster outskirts. Then we should
see a concentration of galaxies around the systemic velocities of NGC
4889 and NGC 4874 only in the cluster center, but not on larger
scales.

The idea that the two subclusters are meeting on an orbit mostly in
the plane of the sky is supported by two further arguments.  First,
the LOS velocity difference between the two supergiant galaxies and
their surrounding cluster cores is only $\simeq 700\kms$, while the
orbital collision velocity must exceed $\sim 2000 \kms$.  To
estimate this we have computed the virial velocity of the Coma cluster
${\rm v}_{200}\simeq 1500\kms$ from the virial masses and radii determined
by \citet{Geller99,Lokas03}. The collision velocity when the cluster
cores meet will be significantly larger than the median subhalo
velocity found by \citet{Hayashi06} from simulations, $1.1{\rm v}_{200}$,
because this is determined for a distribution of subhalo orbital
phases. If we take the relative velocity reached by 10\% of their
subhalos, this gives $1.55{\rm v}_{200}\simeq2300\kms $.   Second,
the Coma cluster is part of the ``Great Wall'' \citep{Geller89}, a
large-scale structure of clusters and filaments that extends
approximately perpendicularly to the line-of-sight. In this structure
Coma is connected through filaments to neighbouring large clusters,
notably A2199, A1367, and A779, at approximate position angles from
North through East of $80\deg$, $250\deg$, and $275\deg$ \citep[see
Fig.~3 of][]{Adami05b}.  E.g., the subcluster around the cD galaxy NGC
4839 appears to be infalling towards the main Coma cluster from the
filament connecting Coma to A1367, at an estimated angle with respect
to the line-of-sight of $74\deg$ and velocity in the plane of the sky
of $\sim 1750\kms$ \citep{Colless96,Neumann01}.

The direction of infall of the NGC 4889 subcluster similarly is likely
to have been from one of the filaments that connect the Coma cluster
with the great wall -- but from which of the three filaments did it
come?  \citet{Biviano96} construct a smoothed map of mean recession
velocity ${\rm v}_{\rm LOS}$ for the fainter galaxies in their Coma
sample. These galaxies are presumably further from the main site of
gravitational interaction near the cluster cores, and thus are better
tracers of the original subcluster motion. Figs.~12, 13 in
\citet{Biviano96} show an interesting gradient in the smoothed 2D
velocity field along a direction $35\deg$ inclined with respect to the
RA axis (i.e., along PA $35\deg$ resp.\ $215\deg$). The mean ${\rm v}_{\rm LOS}$
of the faint galaxies is ${\rm v}_{\rm LOS}\simeq 6700\kms$ at 10'-20' SW of the
center and increases smoothly to above ${\rm v}_{\rm LOS}\simeq 7000\kms$
10'-20' NE of the center. Together with Fig.~\ref{quadranti} we
interpret this gradient in the sense that the NGC 4889 subcluster now
dominates WSW of the center, while the NGC 4874 subcluster is more
prominent ENE of the center. Based on the elongated ICL morphology,
strong interaction has already taken place between the cluster cores,
so that they must have already passed through each other. Then the
observed gradient requires that the NGC 4889 subcluster has last moved
from ENE (the direction of A2199) across the center towards WSW,
whereas the NGC 4874 subcluster must have fallen in from the West (the
direction of A1367 or perhaps A779).

\subsection{Comparison with X-ray data and cluster simulations}

This conclusion finds independent confirmation from the recent X-ray
analysis of \citet{Arnaud01} and \citet{Neumann03}. These authors find
a significant residual X-ray emission in their XMM-Newton data from a
large arc-like region between the Coma cluster core and the NGC 4839
group, and a temperature enhancement on the side of this structure
towards the cluster core. There is also a positive residual on the
south-east side at a similar distance from the center, but with a
cooler temperature; this may be associated with a group around NGC
4911. Hydrodynamic simulations of galaxy cluster collisions
\citep{Roettiger97,Takizawa99,Ricker01,Ritchie02} have shown that the
compression and shocks generated in the collision give rise to large
changes in X-ray luminosity, temperature, and emissivity contour
shapes. After the collision of the cores of two similar mass clusters
a strong arc-shaped shock is driven into the outer parts of the
cluster's X-ray emitting envelope. The morphology of these arcs is
reminiscent of the morphology of the large western X-ray residual in
Coma \citep{Arnaud01,Neumann03}.

While simulations with idealized initial cluster models show strong,
regular, and symmetric features, cluster mergers from cosmological
initial conditions result in much more irregular morphologies
\citep{Rowley04}.  The morphology of their cluster merger 13, about
300 Myr after the collision of the cluster cores, is not unlike that
observed for the Coma cluster. At this time, the emissivity contours
are elongated along the direction of approach and the main arc-shaped
shock is in the forward direction.  From Fig.~3 of \citet{Rowley04},
we estimate that this merger increased the mass of the main cluster by
$\sim 50\%$. Thus a picture in which the NGC 4889 subcluster has
recently fallen in from ENE and its centroid is now past the center of
Coma towards WSW, appears consistent with the presence of the large
western arc in the X-ray emission and the increased X-ray temperature
associated with it. For reference, the western arc shock would have
reached its current radius of $\sim 30$ arcmin from the cluster center
after 300 Myr with an average velocity of $\sim 2700 \kms$.

We mention that our interpretation is different from the suggestion by
\citet{Neumann03} that the western arc could be caused by the hot
plasma of a subcluster that previously merged into Coma from the
southwest, which was heated in the collision and deflected northwards
by the main cluster hot athmosphere. In the hydrodynamic simulations,
such deflections occur when the hot athmospheres collide before the
two cluster cores have met.  This is unlike the situation in the Coma
cluster where, as seen from the southwest, the two subcluster cores
around NGC 4874 and NGC 4889 are ahead of the western arc.  On the
other hand, the NGC 4839 subcluster to the southwest has not yet
reached the main cluster: the X-ray observations \citep{Neumann03}
show a depression in the emission between the NGC 4839 group and main
Coma cluster \citep[see also][]{Neumann01}.

 Our interpretation is also different from the scenario of
\citet{Watanabe99}. Based on an earlier, somewhat different ASCA
temperature map, they suggested that the NGC 4874 subcluster had
merged into the cluster from the southeast. There is no filament in
this direction connecting to a neighbouring cluster. Also the
elongation direction of the ICL distribution and the velocity
differences of the two cD galaxies with respect to their surrounding
galaxy concentrations cannot be explained if both galaxies have just
passed eachother along this direction. 

Cluster mergers can disrupt cooling cores if enough gas from the
secondary cluster can reach the core \citep{Gomez02}, as happens
particularly in head-on collisions of equal-mass clusters
\citep{Ritchie02}.  With the near-head-on geometry of the collision
indicated by the ICL and galaxy velocity measurements, it is likely
that a preexisting cooling core in the cluster would have been heated
by the collision, consistent with the absence of a cooling core in the
Coma cluster now.

\subsection{Second passage of the subcluster cores and predictions
for diffuse light kinematics}
\label{secondpassage}

   \begin{figure}
   \centering
   \includegraphics[width=\hsize,bb=18 184 592 718,clip]{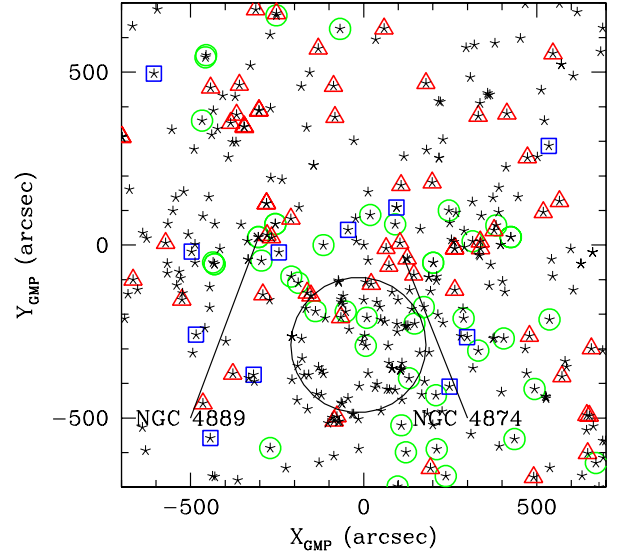}
\caption{ Spatial distribution of all cluster members with available
velocities. The X and Y axes denote RA and Dec, given in arcsec
relative to the GMP83 Coma cluster center, with North up and East to
the left. Green circles are Coma galaxies with $ 6250\kms < {\rm
v}_{\rm LOS} < 6850\kms $, red triangles are galaxies with $ 7050\kms <
{\rm v}_{\rm LOS} < 8000\kms $, blue squares are galaxies with $
4800\kms < {\rm v}_{\rm LOS} < 5350\kms $. The positions of NGC 4874,
NGC 4889 and the MSIS field are indicated on the plot.}
\label{colormap}
\end{figure}

In cosmological simulations of structure formation, substructures
usually collide and merge along highly radial orbits. In a nearly
head-on collision of two extended structures, the outer envelopes of
both structures merge in a slow oscillatory fashion along an
essentially unchanging orbital direction. By contrast, the interaction
of the cores is faster, stronger, and may involve a significant impact
parameter. In this way the cluster cores can have deflected each other
strongly while the motion of the outer clusters still looks apparently
unperturbed \citep[e.g.,][]{Murante07}.

This general picture is the likely explanation for the fact that NGC
4889 is apparently behind (towards the ENE) of the centroid of its
subcluster along the direction of merging as deduced above.  The
elongated distribution of diffuse light requires that the galaxy has
already undergone strong interaction in the core.  Also, the IC star
velocities show that significant diffuse light was lost west of its
current position. All this is best explained if the current direction
of motion of NGC 4889 is towards the east.  I.e., the galaxy has gone
around the other (NGC 4874) subcluster core in a highly eccentric
orbit, moving presumably from north-east to south-west and then in a
sharp turn towards south, east, and again north-east around NGC 4874.
Such an orbit for NGC 4889 and a similar, symmetric orbit for NGC 4874
are sketched in Fig.~\ref{icldistr}.  The diffuse light that we now
see trailing the current position of NGC 4889 was probably lost during
the orbit around the NGC 4874 subcluster core between the first and
second close passages, as indicated on the figure. The material
unbound from both galaxies shortly after their first close passage is
likely to be the origin of the diffuse light lobes in the south-west
and north-east.  Note that little material is unbound from the
colliding galaxies before the first pericentre passage; hence the
absence of ICL with NGC 4874 velocities in our MSIS field.

This predicts that in future observations of diffuse light kinematics
we should see NGC 4889 centered velocities west and south of NGC 4874
and towards the current position of NGC 4889, and more NGC 4874
centered velocities towards the northeast of NGC 4874. The latter
assumes that also NGC 4874 and its cluster core has already been
stripped of significant light, which is plausible in the model above
but less certain considering the more intact nature of the NGC 4874
cluster core as discussed in Section \ref{subcores}.

One may ask whether a trace of the predicted ICL kinematics around the
two cD galaxies can be seen in the velocities of nearby galaxies
dissolved from the cluster cores.  Figure~\ref{colormap} shows the
spatial distribution of galaxies with different velocities in the
central parts of the cluster.  In the distribution of galaxies with
velocities $\pm 300\kms$ relative to the systemic velocity of NGC
4889, there is a hint of a filament connecting this cD galaxy with the
region south of NGC 4874, which is covered by the MSIS field. However,
the noise expected in distributions with such small numbers of galaxies
precludes any more definitive conclusion.

\subsection{The $5000\kms$ peak}
\label{fiveth}

In addition to the dominant peak around NGC 4889's systemic velocity,
the line-of-sight velocity distribution of ICPNe in our MSIS field
contains a secondary (blue) peak at velocities around $5000 \kms$. There
are also three galaxies with velocities around $4500 \kms$ in the
field (Fig.~\ref{fig1}, left and center panels, respectively).  What could
be the origin of these fast-moving objects in the Coma cluster core?

The list of galaxy associations in \citet{Adami05b} contains two groups
with low velocities and within less than 1 Mpc of the MSIS field: G7
(an extended association 0.3-1.2 Mpc east of the center, with 8
members, mean velocity $5614\kms$), and G9 (a compact group of 3
members 0.5 Mpc south-west of the center, with mean velocity
$5710\kms$). Relative to the mean cluster velocity of $\sim 6850\kms$
\citep{Colless96}, these galaxies have velocities of $\sim 1200\kms$
towards the observer, while the ICPNe and galaxies in the blue peak of
the MSIS field velocity distribution have even larger velocities
towards us, $\sim 1600$-$2700\kms$.

It is conceivable that the blue-shifted ICPNe and galaxies in the MSIS
field and some or all of the galaxies in G9 and especially those in G7
are part of a single structure falling through the Coma cluster from
behind. The 8 blue ICPNe would be part of the remains of a galaxy
shredded by the strong tidal field in the dense cluster core,
corresponding to a total stellar mass of a few times $10^9\msun$.
This conversion assumes an $\alpha$ parameter for the first magnitude
in the PN luminosity function of $\sim 10^{-8}$
\citep[e.g.,][]{Aguerri05}, and a mass-to-light ratio of $\sim
5$. Because the MSIS field is located in the dense cluster core, the
fast-moving objects in this field are probably located in a much
deeper part of the Coma potential well than the galaxies in the G7 and
G9 associations. This would explain their significantly faster
line-of-sight velocities relative to the Coma cluster mean.  To
explain a present spread of this structure over $f\mpc$ in the plane
of the sky would also require differences in the velocity component
perpendicular to the line-of-sight, of $\sim 1000 f \kms$ over the
past Gyr. Thus if all the galaxies of G7 and G9 were part of this
structure, it would have had to dissolve early-on during its infall
into the cluster.

\section{Discussion and Conclusions}\label{end}

The measured offset of $\sim 700 \kms$ of the ICPNe associated with
the diffuse light in the \citet{Bernstein95} field, relative to the
systemic velocity of the nearer giant elliptical (cD) galaxy NGC 4874,
indicates that at least part of the diffuse light halo around NGC 4874
is not physically bound to this galaxy.

From the substructure in the galaxy velocities, the disturbed X-ray
emission, and the radio halo in the cluster, it is generally believed
that the Coma cluster is in a state of on-going merging \citep[][and
others]{White93,Colless96,Neumann03,Adami05b}. Here we have shown that
also the LOS velocity distribution of the ICPNe in the center of the
cluster suggests an ongoing merger of two substructures in the Coma
core, which is not yet virialised. The two substructures are
associated with NGC~4889 and NGC 4874.

Starting with the distribution and kinematics of the diffuse light in
the cluster core, we have developed a model for the geometry of this
subcluster merger that is based also on the distribution of Coma
galaxy velocities and the presence of structure in the X-ray
emissivity and temperature distribution. In this model, the NGC 4889
subcluster fell into the region of the Coma cluster from ENE, the
direction of the filament connecting Coma with A2199.  Whereas the
infall direction of the NGC 4874 subcluster was from the west,
probably the direction of the filament towards A1367.  The two
subcluster cores continue to disrupt each other through their mutual
gravitational interaction, with the disruption of the NGC 4889
subcluster core apparently in a more advanced stage. The inclusion of
morphology and kinematics of the ICL in this model was vital because
these data constrain the dynamical evolution in the cluster core
itself.  The ICL kinematics predicted by the model in
further fields can be tested with future observations, as described in
Section~\ref{secondpassage}. 

Currently, the two subcluster cores are past their second, close
passage, during which much of the ICL seen through ICPNe must
have been dissolved. This follows from the absence of any sign
of mixing in the ICPN velocity distribution, and strongly 
supports the conclusion of \citet{Murante07} that a large 
part of the ICL in clusters originates directly or indirectly
from mergers of the most massive galaxies.

Several Gyr in the future, the two supergiant galaxies and their
subcluster cores will have merged. With the further arrival of the NGC
4839 subcluster from the SW, the cluster may not come to rest for
significantly longer.  The part of the galaxy distribution centered
around the NGC 4889 velocity in the 4 quadrant diagram comprises a
significant fraction of all galaxies in the Coma cluster core, perhaps
30\%. This suggests that a truly major subcluster merger is currently
taking place.  Together with the other evidence for infalling
substructures this suggests that Coma is forming now!

\begin{acknowledgements}
We are grateful to the on-site Subaru staff for their support.
We thank C.~Adami and A.~Biviano for making their catalogue
of galaxy velocities available in digital form.
\end{acknowledgements}

\end{document}